\documentstyle[11pt,aaspp4]{article}

\slugcomment{Submitted to The Astronomical Journal}

\lefthead{Schneider et al.}
\righthead{Pair of High Redshift Quasars}

\begin{document}

\title{Discovery of a Close Pair of $z$ = 4.25 Quasars
from the Sloan Digital Sky Survey
\footnote{Based on observations obtained with the Sloan Digital
 Sky Survey, which is owned and operated by the Astrophysical Research
 Consortium.}$^,$\footnote{Based on observations obtained with the Hobby-Eberly
 Telescope, which is a joint project of the University of Texas at Austin,
 the Pennsylvania State University, Stanford University,
 Ludwig-Maximillians-Universit\"at M\"unchen, and Georg-August-Universit\"at
 G\"ottingen.}
}
\author{
Donald~P.~Schneider\altaffilmark{\ref{PennState}},
Xiaohui~Fan\altaffilmark{\ref{Princeton}},
Michael~A.~Strauss\altaffilmark{\ref{Princeton}},
James~E.~Gunn\altaffilmark{\ref{Princeton}},
Gordon~T.~Richards\altaffilmark{\ref{PennState}}$^,$\altaffilmark{\ref{Chicago}},
G.R.~Knapp\altaffilmark{\ref{Princeton}},
Robert~H.~Lupton\altaffilmark{\ref{Princeton}},
David~H.~Saxe\altaffilmark{\ref{IAS}},
John~E.~Anderson,~Jr.\altaffilmark{\ref{FNAL}},
Neta~A.~Bahcall\altaffilmark{\ref{Princeton}},
J.~Brinkmann\altaffilmark{\ref{APO}},
Robert~Brunner\altaffilmark{\ref{Caltech}},
Istvan~Csab\'ai\altaffilmark{\ref{JHU}}$^,$\altaffilmark{\ref{Hungary}},
Masataka~Fukugita\altaffilmark{\ref{CosJapan}}$^,$\altaffilmark{\ref{IAS}},
G.S.~Hennessy\altaffilmark{\ref{USNODC}},
Robert~B.~Hindsley\altaffilmark{\ref{NRL}},
\v Zeljko~Ivezi\'c\altaffilmark{\ref{Princeton}},
R.C.~Nichol\altaffilmark{\ref{CMU}},
Jeffrey~R.~Pier\altaffilmark{\ref{USNOAZ}},
and
Donald~G.~York\altaffilmark{\ref{Chicago}}
}

email addresses: dps@astro.psu.edu, fan@astro.princeton.edu,
strauss@astro.princeton.edu

\newcounter{address}
\setcounter{address}{3}
\altaffiltext{\theaddress}{Department of Astronomy and Astrophysics, The
   Pennsylvania State University, University Park, PA 16802.
\label{PennState}}
\addtocounter{address}{1}
\altaffiltext{\theaddress}{Princeton University Observatory, Princeton,
   NJ 08544.
\label{Princeton}}
\addtocounter{address}{1}
\altaffiltext{\theaddress}{Astronomy and Astrophysics Center, University of
   Chicago, 5640 South Ellis Avenue, Chicago, IL 60637.
\label{Chicago}}
\addtocounter{address}{1}
\altaffiltext{\theaddress}{The Institute for Advanced Study, Princeton,
   NJ 08540.
\label{IAS}}
\addtocounter{address}{1}
\altaffiltext{\theaddress}{Fermi National Accelerator Laboratory, P.O. Box 500,
   Batavia, IL 60510.
\label{FNAL}}
\addtocounter{address}{1}
\altaffiltext{\theaddress}{Apache Point Observatory, P.O. Box 59,
   Sunspot, NM 88349-0059.
\label{APO}}
\addtocounter{address}{1}
\altaffiltext{\theaddress}{Astronomy Department, California Institute of
   Technology, Pasadena, CA 91125.
\label{Caltech}}
\addtocounter{address}{1}
\altaffiltext{\theaddress}{Department of Physics and Astronomy,
   Johns Hopkins University, 3701 University Drive, Baltimore, MD 21218.
\label{JHU}}
\addtocounter{address}{1}
\altaffiltext{\theaddress}{Department of Physics of Complex Systems,
   E\"otv\"os University, P\'azm\'ay P\'eter \hbox{s\'et\'any 1/A,}
   H-1117, Budapest, Hungary.
\label{Hungary}}
\addtocounter{address}{1}
\altaffiltext{\theaddress}{Institute for Cosmic Ray Research, University
   of Tokyo, Midori, Tanashi, Tokyo 188-8588, Japan
\label{CosJapan}}
\addtocounter{address}{1}
\altaffiltext{\theaddress}{US Naval Observatory, 3450 Massachusetts Avenue NW,
   Washington, DC 20392-5420.
\label{USNODC}}
\addtocounter{address}{1}
\altaffiltext{\theaddress}{Remote Sensing Division, Code~7215, Naval Research
   Laboratory, 4555~Overlook~Ave. SW, Washington, DC~20375.
\label{NRL}}
\addtocounter{address}{1}
\altaffiltext{\theaddress}{Dept. of Physics, Carnegie Mellon University,
     5000~Forbes Ave., Pittsburgh, PA~15232.
\label{CMU}}
\addtocounter{address}{1}
\altaffiltext{\theaddress}{US Naval Observatory, Flagstaff Station,
   P.O. Box 1149, Flagstaff, AZ 86002-1149.
\label{USNOAZ}}

\vbox{
\begin{abstract}
We report the discovery of a pair of $z = 4.25$ quasars with a separation
of~33$''$.  The brighter of the two objects was identified as a high-redshift
quasar candidate from Sloan Digital Sky Survey multicolor imaging data, and
the redshift was measured from a spectrum obtained with the Hobby-Eberly
Telescope.  The slit orientation of this observation {\it by chance}
included another quasar, approximately
one magnitude fainter and having the same redshift as the target.  This is
the third serendipitous discovery of a \hbox{$z > 4$} quasar.
The differences in the relative strengths and profiles of the emission
lines suggest that this is a quasar pair and not a gravitational lens.
The two objects are likely to be physically associated; the
projected physical separation is approximately 210~$h_{50}^{-1}$~kpc and the
redshifts are identical to~$\approx$~0.01, implying a radial
physical separation of 950~$h_{50}^{-1}$~kpc or less.
The existence of this pair is strong circumstantial evidence
\hbox{that $z \sim 4$} quasars are clustered.

\end{abstract}
}



%

\section{Introduction}

One of the first observations of the $z > 4$ universe was also one of the
most startling: the serendipitous discovery of the \hbox{$z = 4.4$}
quasar \hbox{Q 2203+29} by McCarthy~et~al.~(1988).  While obtaining a spectrum
of the \hbox{$z = 0.707$} radio galaxy 3C~441, the spectrograph slit was
aligned along the radio jet.  The data included the spectrum of a faint
red source~51$''$ from the nucleus; the 
secondary source's redshift of~4.41 made it
the second most distant known object at the time!
The authors showed that this was a very low probability event unless
the number density of high-redshift quasars had been underestimated by
orders of magnitude (a slight possibility at the time given that only a few
\hbox{$z > 4$} quasars were known and thus the number density
poorly constrained).

Several years later \hbox{PC 0027+0521}, a faint $z=4.21$ quasar,
was found by chance in a high-redshift quasar survey (Schneider, Schmidt,
\& Gunn~1994).  The object happened to lie in the slit
($i.e.,$ the slit was placed at the parallactic angle, not rotated to
acquire additional objects) while the spectrum of a
high-redshift quasar candidate (which turned out to be a star!) was obtained.

{\it Post facto} calculations showed that the chance of either event occurring
was on the order of~0.1\%.  In this paper we describe the third serendipitous
discovery of \hbox{a $z > 4$ quasar}; it was discovered, like
\hbox{PC 0027+0521}, when by chance it was included in a spectroscopic
observation of a high-redshift quasar candidate.  In this case, however,
the candidate was indeed a quasar, and it has the same redshift as the
serendipitous source.

\section{Observations}

\subsection{Sloan Digital Sky Survey} \label{survtech}

The Sloan Digital Sky Survey (SDSS; \cite{york00})
uses a CCD camera (\cite{gcam98}) on a
dedicated 2.5-m telescope (\cite{ws01}) at Apache Point Observatory,
New Mexico, to obtain images in five broad optical bands over
10,000~deg$^2$ of the high Galactic latitude sky centered approximately
on the North Galactic Pole.  The five filters (designated $u'$, $g'$,
$r'$, $i'$, and $z'$) cover the entire wavelength range of the CCD
response (\cite{fig96}).  Photometric calibration is provided by simultaneous
observations with a 20-inch telescope at the same site.  The
survey data processing software measures the properties of each detected object
in the imaging data, and determines and applies astrometric and photometric
calibrations
(\cite{jrp01,rhl01}).

The high photometric accuracy of the SDSS images and the information
provided by the $z'$ filter (central wavelength of 9130~\AA ) make the
SDSS data an excellent source for identification of high-redshift
quasar candidates.  In the past two years the SDSS has identified
\hbox{many $z > 3.5$} quasars (Fan~et~al.~1999a, 2000a,c;
Schneider~et~al.~2000; Zheng~et~al.~2000),
culminating with the discovery of \hbox{a $z = 5.80$} quasar (\cite{fan00b}).

We have started a survey of faint, high-redshift quasars using the SDSS
imaging data.  This survey uses a multicolor selection technique
similar to that of \hbox{Fan et al. (1999a,2000a)};
the major modification is a change from $i^*$~=~20.0
to $i^*=21.0$ for the limiting magnitude.  \hbox{The $i^* \approx 20.6$} object
\hbox{SDSSp J143952.58$-$003359.2} was flagged as a probable
\hbox{$z \approx 4.3$} quasar, primarily based on its location in
the \hbox{($g^*- r^*$), ($r^* - i^*$)} diagram, from imaging data taken
on 22~March~1999 (SDSS imaging run~756).
For brevity, we will sometimes refer to this
object as ``A" in this paper.

The area containing the quasar
had also been observed on 20~March~1999 (SDSS imaging run~745).
The SDSS photometric measurements of the quasar on both nights
are presented in Table~1, which shows that the photometry/errors are
internally consistent.

Notes on SDSS nomenclature:  The names for sources have the format
\hbox{SDSSp Jhhmmss.ss+ddmmss.s}, where the coordinate
equinox is~J2000, and the ``p" refers to preliminary.  The reported magnitudes
are based on a preliminary photometric calibration; to indicate this
the filters have an asterisk instead of a prime superscript.  The
estimated astrometric accuracies in each coordinate are~0.1$''$ and the
calibration of the photometric measurements are accurate to~0.05~magnitudes.
Note that the $u^*$ measurements indicate essentially zero flux in this band.

\subsection{Spectroscopy of Quasar Candidates} \label{survfields}

A spectrum of \hbox{SDSSp J143952.58$-$003359.2} was obtained on
\hbox{2 June 2000} with the Marcario
Low Resolution Spectrograph (LRS; Hill~et~al.~1998a,1998b)
of the Hobby-Eberly Telescope (HET; \cite {lwr98})
\footnote{A summary of the
commissioning of the LRS is given in Schneider~et~al.~(2000).}.
The LRS is mounted in the Prime Focus Instrument Package, which
rides on the HET tracker.  The 900~s observation was acquired with the~2$''$
wide slit, rotated to the parallactic angle.
The dispersive
element was \hbox{a 300 line mm$^{-1}$} grism blazed \hbox{at 5500 \AA .}
An OG515 blocking filter was installed to permit calibration of the spectra
out to 1~micron.
The detector is a thinned, antireflection-coated
3072~$\times$~1024 Ford Aerospace CCD, and was
\hbox{binned $ 2 \times 2$}
during readout; this produced
an image scale of~0.50$''$~pixel$^{-1}$ and a dispersion
\hbox{of $\approx$ 4.5 \AA\ pixel$^{-1}$}.  The spectra covered the
range from 5100--10,200~\AA\ at a resolution of approximately~20~\AA.

The wavelength calibration was provided by Ne, Cd, and Ar comparison lamps, and
the relative flux and atmospheric band absorption calibration was provided by
the observation of the spectrophotometric standard \hbox{BD +26$^{\circ}$ 2606}
(Oke \& Gunn~1983).  The observing conditions were marginal; the seeing
was 2$''$ with moderate cloud cover (the signal from the standard indicated
more than a magnitude of extinction).
Absolute
spectrophotometric calibration was carried out
by scaling each spectrum so that
$i^*$ magnitudes synthesized from the spectra matched the SDSS photometric
measurements.

Figure~1 shows the SDSS $i'$ image of the field of
\hbox{SDSSp J143952.58$-$003359.2}, which is denoted by an ``A" in the figure;
this object was placed in the center of the slit.  The slit was oriented
at a position angle of~26$^{\circ}$; the spectra of the
bright star and object ``B", both located
southwest of A, were obtained as they fell into the slit.  A single 900~s
exposure was made.

Figure~2 displays the region centered at~6680~\AA\ of the spectroscopic
exposure.
The spectra run vertically from 5790~\AA\ (bottom) to 7575~\AA \ (top).
The data are dominated by the spectrum of the star
slightly to the right of center.  The spectrum of~A is clearly visible
in the center of the frame, and the spectrum possesses the characteristic
signal of a high-redshift quasar (strong, broad emission line, and a
significant drop in the continuum as one moves from red to blue across the
line).  The initial visual inspection of the frame revealed that the spectrum
of~B, located midway between the star and the right of the frame in Figure~2,
appeared to have a very similar spectrum to~A.

The flux and wavelength calibrated
spectra of the two objects are presented in Figure~3 (the brighter object,
A, is the top panel).  Although the signal-to-noise ratio is not large
(6 per binned pixel at 7000~\AA\ in~A, a factor of two lower in~B),
there is little doubt
that both objects \hbox{are $z \approx 4.25$} quasars: the broad,
strong, asymmetric Lyman~$\alpha$ emission line and the continuum depression
due to Lyman~$\alpha$ forest absorption are the most obvious features.
The Lyman~$\alpha$ forest absorption appears to be significantly
different in the two quasars; both the photometry and the spectroscopy indicate
that the spectrum of~A suffers considerably less absorption than~B.
We are unable to search for any Lyman limit absorption because the
OG515 filter does not transmit light below a
wavelength of~$\approx$~980~\AA\ in
the rest frame of the quasars.

The redshifts, determined from the Si+O and C~IV emission lines in~A and
the C~IV line in~B, are identical
to within the measurement errors.   The two objects, however, are
almost certainly a pair of quasars and not a lens, given the
relatively large separation and markedly different profiles
of the Lyman~$\alpha$ emission line and the contrast in the strengths of
the~Si+O feature.

\section{Discussion}

Object ``B" was identified with \hbox{SDSSp J143951.60$-$003429.2}
in the SDSS database; its photometric measurements are presented in
Table~1.  
Figure~4 shows the locations of~B and~A in the
\hbox{$g^*-r^*$,$r^*-i^*$} diagram, which is the SDSS color-color plot
in \hbox{which $z \approx 4$} are most clearly separated from
the stellar locus (the Lyman~$\alpha$ emission
line falls in~$r'$, the Lyman forest occurs in~$g'$; see Fan~1999).
Object~A satisfies the high-redshift selection criteria in this diagram
(shaded area) and is approximately~0.4 magnitudes brighter than
\hbox{the $i^* = 21.0$} limit.  Object B~lies in the same direction as~A
from the stellar locus, but is slightly outside the selection region and
is also nearly half a magnitude too faint in~$i^*$~to be included in the
sample.  For both objects the $g'$ measurement has a very low
signal-to-noise ratio.

A summary of the properties of the two quasars is given in Table~2;
the absolute magnitudes were calculated assuming
\hbox{$H_0$ = 50 km s$^{-1}$ Mpc$^{-1}$} \hbox{and $q_0$ = 0.5.}
(We will use this cosmological model throughout this section.)
We do not quote a continuum depression value ($D_A$) because of the
limited signal-to-noise ratio of the data.
The quasars are relatively faint (indeed, the fainter of the pair is
quite similar in redshift and luminosity to the second serendipitous
\hbox{$z > 4$ object,} \hbox{PC 0027+0521}), but the luminosity of each
is well above the canonical ``quasar/Seyfert" dividing line
\hbox{of $M_B = -23$.}  
The observed radio emission from each quasar must be
less than 1~mJy based on the lack of detection in the
FIRST (Becker, White \& Helfand~1995) survey data;
the quasars are radio-quiet, as are the vast majority of \hbox{$z > 4$ quasars}
(Schneider et al.~1992; Schmidt et al.~1995; Stern et al.~2000;
Carilli~et~al.~2000).

The quasars are separated by~33.4$''$ and~B is at a position angle
of~206$^{\circ}$ relative to~A.  
In our adopted cosmology, this corresponds to a projected spatial separation
of~208~kpc (1.1~Mpc comoving), much closer than any other published pair
of \hbox{$z > 4$} quasars.
We have measured the
redshift difference in the spectra both from individual line redshifts and
cross-correlating the spectra; both techniques yield a redshift difference
consistent with zero but with an uncertainty of~0.01.

If the redshift difference is~0.01 and this is due to cosmological expansion
rather than peculiar velocities,
then the radial separation is~950~kpc (5~Mpc comoving).
It is quite possible, however,
that the redshift difference is much smaller, in which case assigning
a distance to the redshift difference is suspect.  If we assume that
we have no radial information and are viewing the pair at a random angle,
the mean deprojected separation is 327~kpc (1.7~Mpc comoving).

The only other known pair of $z > 4$ quasars with a comoving separation of less
than~100~Mpc was found in observations for
the Palomar Scan Grism Survey (PSGS; Schneider, Schmidt, \& Gunn 1999):
\hbox{PC 0027+0525} ($z = 4.099$, $M_B = -24.8$) and the previously mentioned
serendipitously discovered \hbox{PC 0027+0521} ($z = 4.210$, $M_B = -24.0$).
This pair is separated on the sky by~319$''$ and has a
comoving separation of approximately 58~Mpc.

Both these pairs are chance discoveries, so calculation of the probability
of finding them will necessarily involve
{\it post facto} arguments.  With this caveat in mind, we investigate this
probability, first assuming that quasars are unclustered, then taking
clustering into account.
What are the chances of finding pairs of quasars
of a given spatial separation between redshifts of 4.0 and~4.5 where both
members \hbox{have $M_B < -24.0$?}
In the following discussion, we will examine three examples: a ``minimum"
separation model (330~kpc; 1.7~Mpc comoving) and a ``maximum" separation model
(1~Mpc; 5.3~Mpc comoving) for the
SDSS pair, and a 58~Mpc separation for the PSGS objects.

Fan~et~al.~(2000d) have investigated the luminosity function of
high-redshift quasars from a sample of~39 \hbox{$z > 3.6$}
objects identified by the SDSS (18 quasars with redshifts larger than four).
They find the comoving number density of 
quasars more luminous than \hbox{$M_B \le -26$} at a redshift of~4.25 to be
$2.8 \times 10^{-8}$~Mpc$^{-3}$, and
\hbox{an $n \propto L^{-2.5}$} differential luminosity
function.  This numbers are in excellent agreement with the previous
studies by
Schmidt, Schneider, \& Gunn~(1995) and Kennefick, Djorgovski, \&
de~Carvalho~(1995), which were based on samples that contained 9 and~10
\hbox{$z > 4$} quasars, respectively.
This luminosity function produces a number density 
\hbox{of $4.4 \times 10^{-7}$ Mpc$^{-3}$}
for quasars with $M_B < -24.0$.  This calculation requires an extrapolation
of more than 1.5~magnitudes from the Schmidt~et~al. and Fan~et~al. data,
and should be viewed as an upper limit
on the number density given the evidence ($e.g.,$ the PSGS) that the
quasar luminosity function flattens at luminosities \hbox{below $M_B = -26$.}

For the three examples mentioned above, the volumes enclosed by spheres whose
radii are equal to the separations are 20.6~Mpc$^3$, 624~Mpc$^3$,
\hbox{and $8.2 \times 10^5$ Mpc$^3$};
the chances of more than one quasar occurring in each volume
are $4.1 \times 10^{-11}$, $3.9 \times 10^{-8}$, and~0.051.
The volume enclosed by the shell
bounded by redshifts of~4.0 and~4.5 is~$1.4 \times 10^{11}$~Mpc$^3$
so one would expect~0.28, 8.9, and~8700 such pairs over the entire sky.
Given that the PSGS/SDSS have examined only~1\% of the celestial sphere
for high-redshift quasars, the likelihood of finding a 1.7~Mpc pair is 
approximately one part in~350 and there is only a nine percent chance of
identifying a 5.3~Mpc pair.  The existence of only two, not~87, known pairs
with smaller than 58~Mpc separation is merely a reflection of the
fact that the current sample
of $z > 4$ quasars is woefully incomplete \hbox{to $M_B = -24$,} which
comes as no surprise.  This implied incompleteness, much more than a factor
of ten, makes the discovery of the SDSS pair even more striking.

This calculation suggests that given the existence of the SDSS pair, it
is very unlikely that high-redshift quasars are not clustered.
Kundi\'c~(1997) and Stephens~et~al.~(1997) found that luminous $z > 2.7$ quasars
are clustered based on the three pairs with less than~30~Mpc (comoving)
separation found in the
90 high-redshift quasars of the Palomar Grism Transit Survey (Schneider,
Schmidt, and Gunn~1994).  Their estimate of the comoving correlation length
was~35~Mpc, with a relatively large uncertainty.  Bahcall \& Chokshi~(1991)
discussed the origin of low-redshift quasar correlations, and
two recent papers
(Martini \& Weinberg~2000; Haiman \& Hui~2000) predict that high-redshift
quasars should be clustered based on theoretical arguments.
\smallskip

\vbox{
How strong must the quasar-quasar correlation function be to produce the
observed SDSS
pair in only~1\% of the sky?  If we assume a correlation function of the
form seen in the local universe and that for \hbox{$z \approx 3$}
galaxies (Giavalisco~et~al.~1998)

$$ \xi(r) \ = \ \left({r \over r_0}\right)^{-\gamma} $$

\noindent
the ratio of the number of objects in the clustered case to the unclustered
case within a sphere of radius~$R$ is

$$ {n_{\rm corr} \over n_{\rm field}}
\bigg | _R \ = \ {\int_0^R 4 \pi r^2 \xi(r)
\ dr \over \int_0^R 4 \pi r^2 \ dr} \ = \ {3 \over 3 - \gamma}
\
\left({R \over r_0}\right)^{-\gamma} $$
}

\smallskip
\vbox{\noindent
Adopting a representative value of $\gamma$~=$1.8$, this expression
becomes

$$ {n_{\rm corr} \over n_{\rm field}} \bigg | _R \ =
\ 2.5 \left({R \over r_0}\right)^{-1.8} $$
}

If this expression describes the high-redshift quasar-quasar correlation
function, then the scale length~$r_0$
must approach 30~Mpc for the minimum model and
12~Mpc for the maximum model for there to be a reasonable chance
of finding a pair in the SDSS database to date.  (Remember that
this calculation
assumes that the quasar sample is complete to \hbox{$M_B = -24$!)}

One could argue that the above discussion is invalid because
current surveys are failing to find most of the quasars
at this redshift due to effects such as reddening
($e.g.,$ Ostriker \& Heisler~1984); therefore the estimate of
the mean number density of quasars is grossly underestimated and
hence ``chance" pairs much more prevalent the calculation suggests.
This may
indeed be the state of affairs, but even if true
would not invalidate the above reasoning.  The
quasar surveys mentioned above have detection algorithms that rely on the
strength of the Lyman~$\alpha$ absorption and strong Lyman~$\alpha$ emission
(although the SDSS has demonstrated the ability to find high-redshift objects
without Lyman~$\alpha$ emission; see Fan et al.~1999b).  Perhaps we can
identify
but a tiny fraction of the luminous objects at redshift greater than four,
but we can state that
luminous objects with strong Lyman~$\alpha$ emission are clearly
strongly clustered.

These numbers indicate that it is highly likely that a significant number
of $z >4$ quasar pairs will be discovered in current large area surveys.
Recent investigations of the environments of high-redshift quasars
($e.g.,$~Djorgovski~1999) demonstrate that these distant beacons
do not occur in isolation.  The SDSS spectroscopic survey
should discover several thousand
quasars with redshifts larger than four, and the SDSS image database
will contain useful color information for objects brighter than
\hbox{$i' \approx 21.5$}, similar to the brightness of the fainter of the
SDSS pair in this study, in all of the fields.  While the efficiency of 
identification of high-redshift quasars in the primary SDSS survey
at this limit will fall well below the \hbox{50\%--70\%} levels seen
in the SDSS to date ($e.g.,$ \hbox{Fan et al. 1999a,2000a)},
the number of such candidates within an
arcminute or two of
known high-redshift quasars should be sufficiently small for spectroscopic
investigation.  The southern SDSS survey, consisting of an area of~225~sq~deg
that will reach a magnitude or more deeper than the primary SDSS survey,
should be ideal for studies of the high-redshift quasar correlation function.

\acknowledgments

We would like to thank Gary Hill, Marsha Wolf, Grant Hill, and Matt
Shetrone for aid with the HET data acquisition, and John Booth and his team
for their efforts that have markedly improved the HET's image quality
during the month preceding the observation.

The Sloan Digital Sky Survey (SDSS) is a joint project of the University of
Chicago, Fermilab, the Institute for Advanced Study, the Japan Participation
Group, the Johns Hopkins University, the Max-Planck-Institute for Astronomy,
Princeton University, the United States Naval Observatory, and the University
of Washington.  Apache Point Observatory, site of the SDSS, is operated by
the Astrophysical Research Consortium.  Funding for the project has been
provided by the Alfred P.~Sloan Foundation, the SDSS member institutions,
the National Aeronautics and Space Administration, the National Science
Foundation, the U.S.~Department of Energy, and Monbusho.
The SDSS Web site \hbox{is {\tt http://www.sdss.org/}.}

The Hobby-Eberly Telescope (HET) is a joint project of the University of Texas
at Austin,
the Pennsylvania State University,  Stanford University,
Ludwig-Maximillians-Universit\"at M\"unchen, and Georg-August-Universit\"at
G\"ottingen.  The HET is named in honor of its principal benefactors,
William P. Hobby and Robert E. Eberly.  
The LRS is named for Mike Marcario of High Lonesome Optics who fabricated 
several optics for the instrument but died before its completion.

This work was supported in part by National Science Foundation grants
AST99-00703~(DPS) and AST96-18503~(MAS).
MAS and XF acknowledge
additional support from Research Corporation, the Princeton University
Research Board, and a Porter O.~Jacobus Fellowship.

\clearpage

%
%

%
%
\newpage

\begin{figure}
\plotfiddle{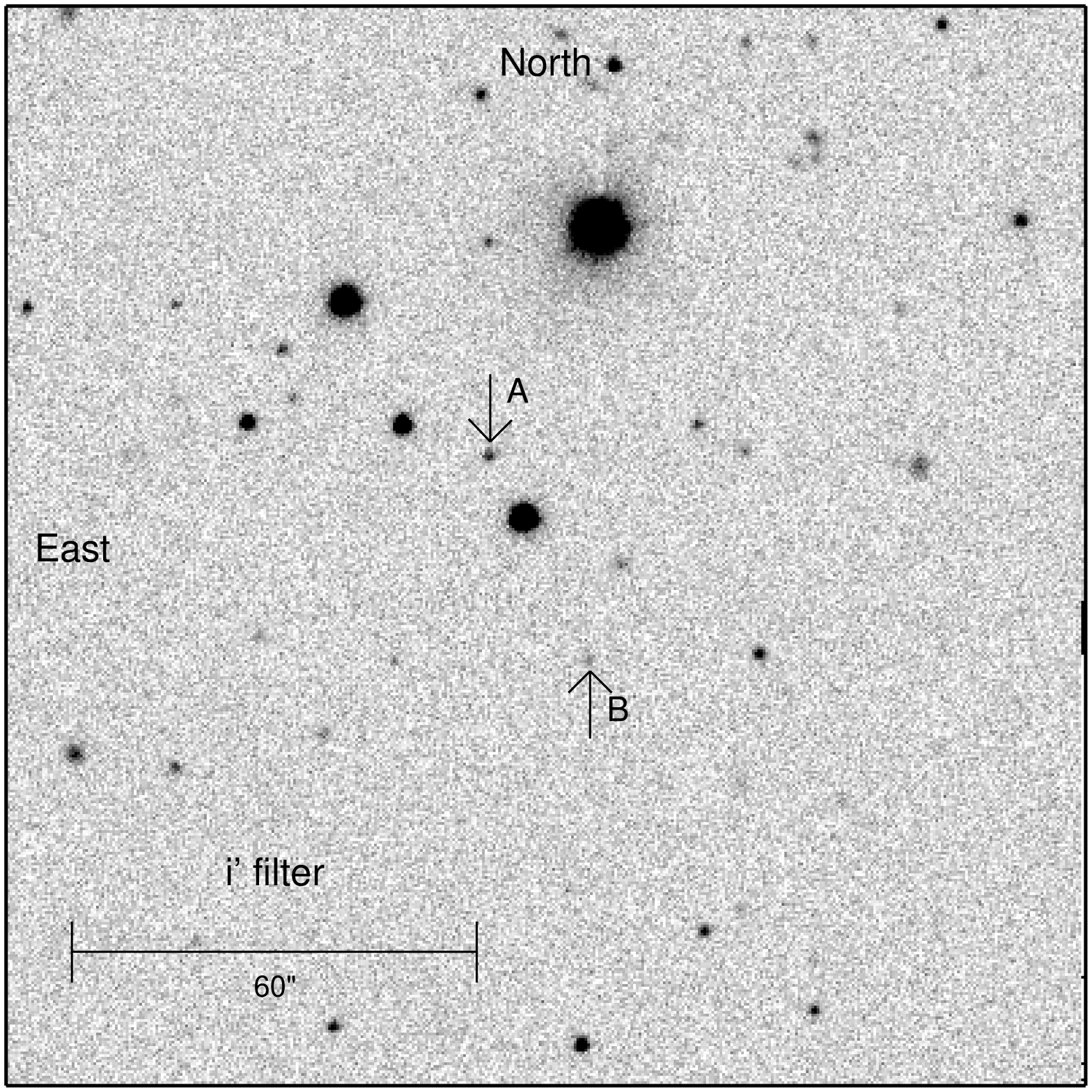}{6.5in}{0.0}{90.0}{90.0}{-255.0}{-32.0}
\figcaption{
Finding chart for the field; north is up and east to the left, and the
chart is~160$''$ on a side.  This is the $i'$ image taken with the SDSS
camera.  The brighter quasar (the target of the spectroscopic observation),
\hbox{SDSSp J143952.58$-$003359.2,}
is denoted by the letter A; B identifies the fainter quasar.  The
HET/LRS observation had the slit oriented at the parallactic
angle and included spectra of A, B, and the star between the quasars.
\label{fig1}}
\end{figure}

\begin{figure}
\plotfiddle{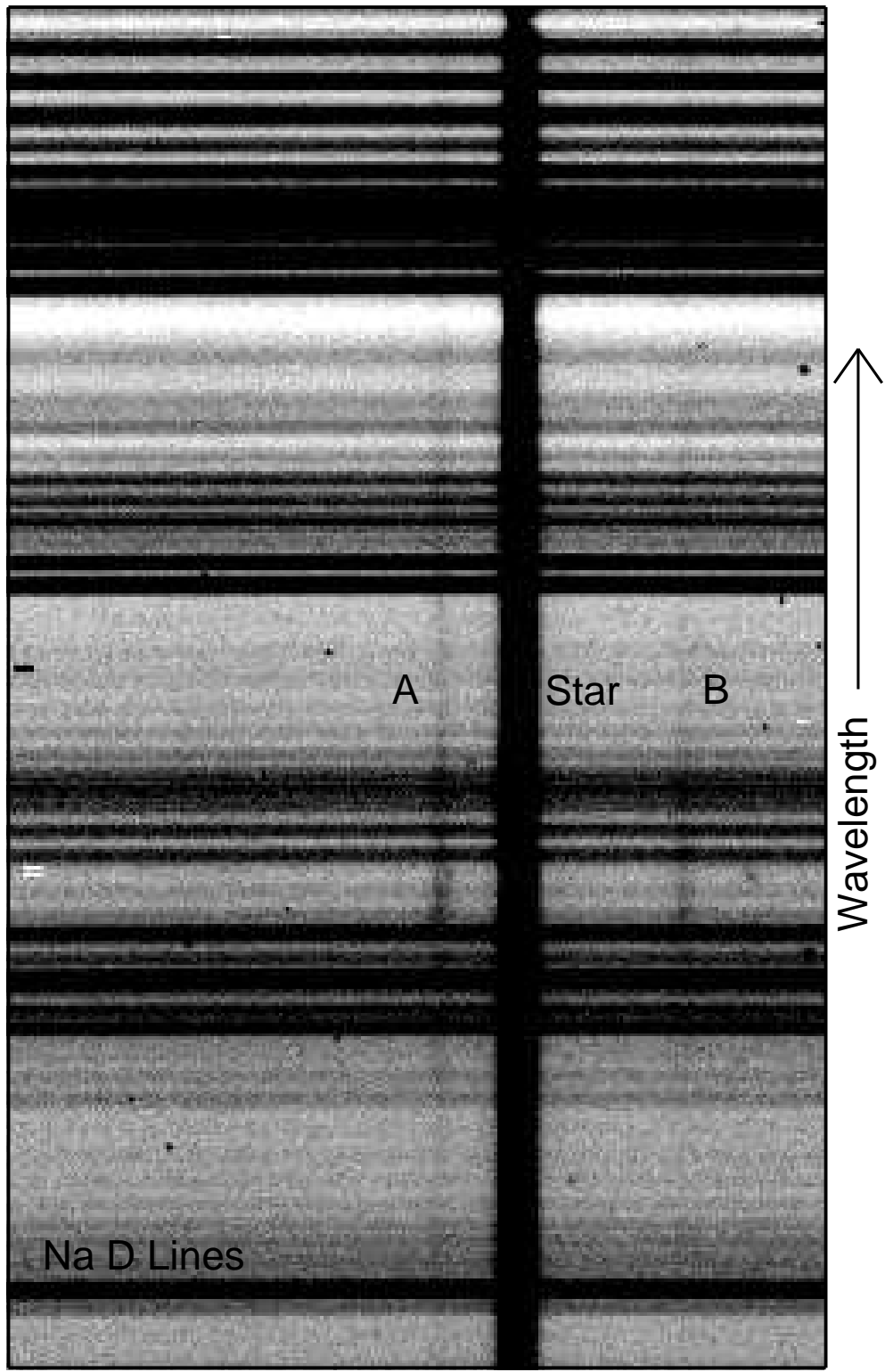}{7.0in}{0.0}{80.0}{80.0}{-180.0}{-30.0}
\figcaption{
This is a section of the flattened LRS frame that contains the Lyman~$\alpha$
emission lines of the two quasars.
Blue is towards the bottom of the frame (the bright skyline
near the bottom is Na~D) and the slit width is 2$''$ (18~\AA).  The width
of the figure is 125$''$.
The spectrum of the brighter quasar (A) is in the center
of the frame, the star's spectrum is just to the right of this, and the spectrum
of the fainter quasar (B)
lies approximately midway between the star's spectrum
and the right edge of the figure.
\label{fig2}}
\end{figure}

\begin{figure}
\plotfiddle{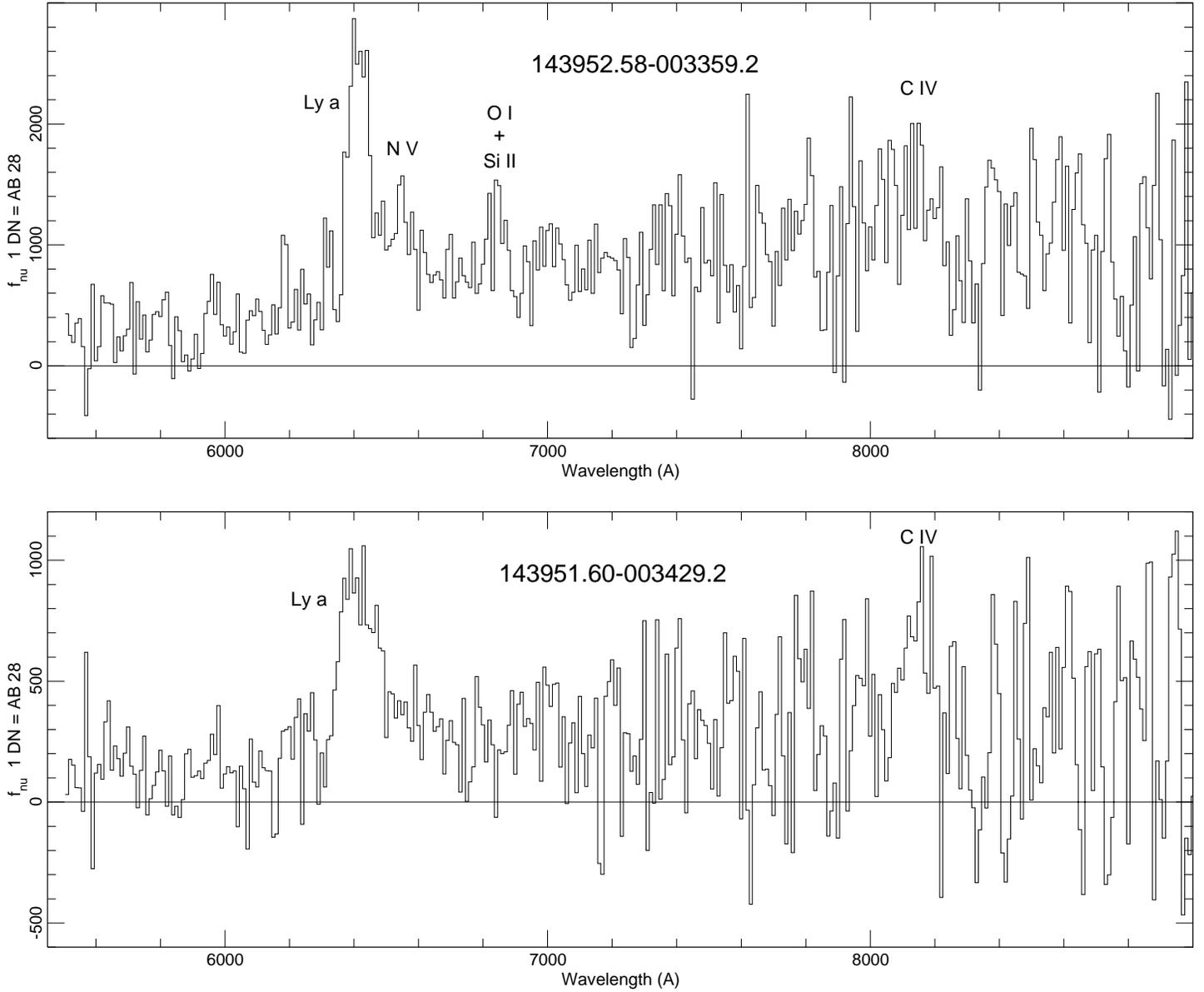}{5.0in}{90.0}{80.0}{80.0}{345.0}{-15.0}
\figcaption{
Spectra of the two quasars taken with the Low-Resolution Spectrograph
on the Hobby-Eberly Telescope.  The exposure time was 900~s; the data were
taken through moderate cloud cover.  The data have been rebinned
\hbox{to 10~\AA\ pixel$^{-1}$.}
The spectral resolution is 20~\AA ;
the unit of flux is \hbox{AB = 28.0} or
\hbox{$2.29 \times 10^{-31}$ erg cm$^{-2}$ s$^{-1}$ Hz$^{-1}$.}  Prominent
emission lines are marked.
\label{fig3}}
\end{figure}

\begin{figure}
\plotfiddle{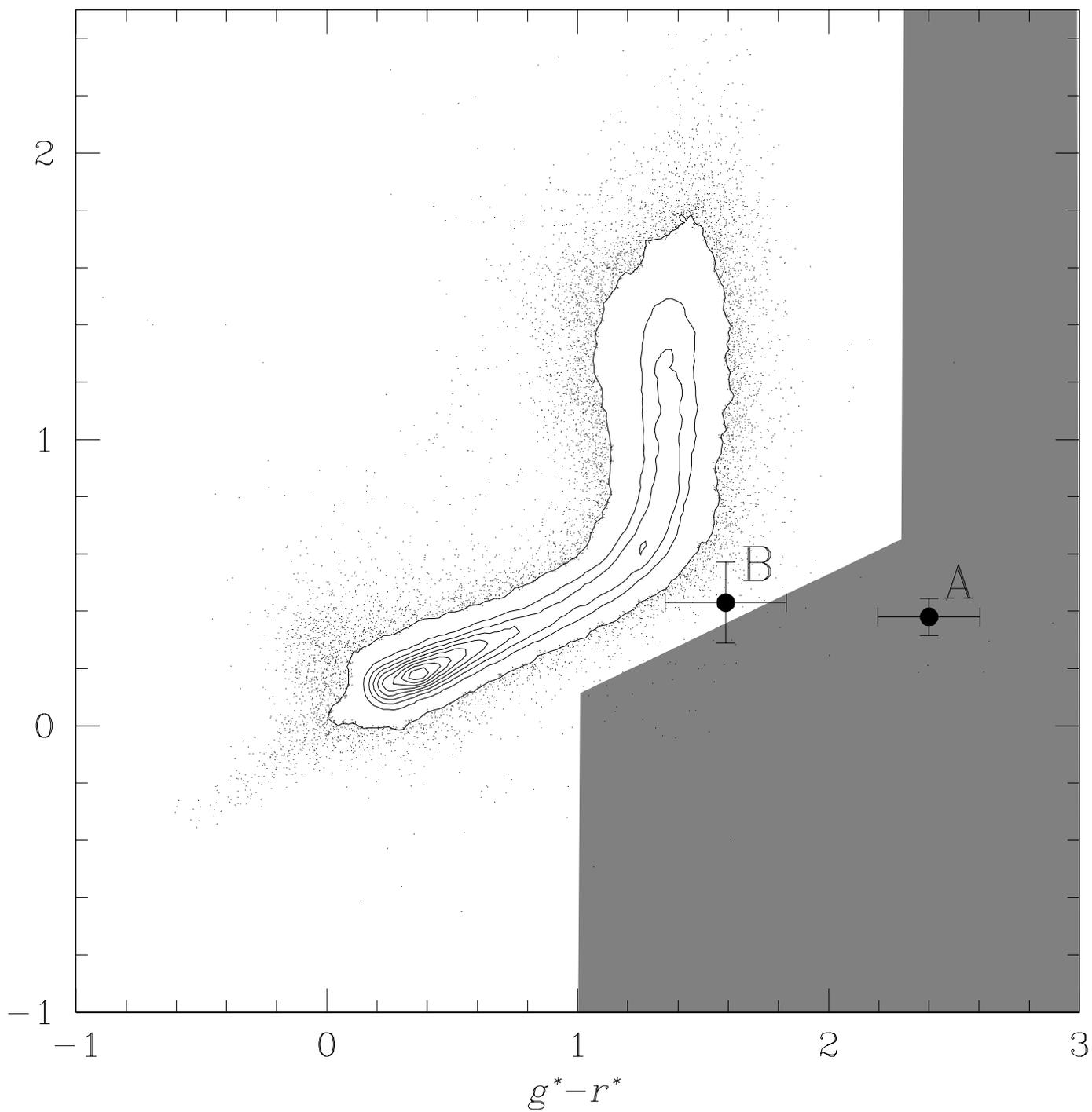}{7.0in}{0.0}{100.0}{100.0}{-295.0}{-142.0}
\figcaption{
Location of the two quasars in the \hbox{$g^*-r^*$,$r^*-i^*$}
diagram.  The shaded area is the selection region for high-redshift
quasars. The points/contours indicate the distribution of stars brighter
than \hbox{$i^* = 21$} selected from an area of 25~square~degrees.
\label{fig4}}
\end{figure}
\clearpage

\halign{\hskip 12pt
# \hfil \tabskip=1em plus1em minus1em&
\hfil # \hfil &
\hfil # &
\hfil # &
\hfil # &
\hfil # &
\hfil # \cr
\multispan7{\hfil TABLE 1. SDSS Photometry$^a$ \hfil}\cr
\noalign{\bigskip\hrule\smallskip\hrule\medskip}
& \hfil SDSS \hfil \cr
\hfil Object$^b$ \hfil&\hfil Run \hfil &\hfil $u^*$ \hfil&\hfil $g^*$ \hfil &
\hfil $r^*$ \hfil & \hfil $i^*$ \hfil & \hfil $z^*$ \hfil \cr
\noalign{\medskip\hrule\bigskip}
SDSSp J143952.58$-$003359.2 & 745 & 23.17 & 23.41 & 20.98 & 20.56 & 20.20 \cr
&& $\pm$0.51 & $\pm$0.24 & $\pm$0.05 & $\pm$0.06 & $\pm$0.16 \cr
SDSSp J143952.58$-$003359.2 & 756 & 24.20 & 23.35 & 20.95 & 20.57 & 20.27 \cr
&& $\pm$0.59 & $\pm$0.20 & $\pm$0.04 & $\pm$0.05 & $\pm$0.12 \cr
\noalign{\medskip}
SDSSp J143951.60$-$003429.2 & 745 & 23.58 & 23.11 & 21.69 & 21.67 & 21.39 \cr
&& $\pm$0.61 & $\pm$0.19 & $\pm$0.09 & $\pm$0.15 & $\pm$0.42 \cr
SDSSp J143951.60$-$003429.2 & 756 & 22.95 & 23.48 & 21.89 & 21.46 & 20.95 \cr
&& $\pm$0.38 & $\pm$0.22 & $\pm$0.10 & $\pm$0.10 & $\pm$0.21 \cr
\noalign{\smallskip}
\noalign{\medskip\hrule}}
\medskip\noindent
$^a$ Photometry is reported in terms of asinh magnitudes; see 
Lupton, Gunn, \& Szalay~(1999) for details.  In this system, zero flux
corresponds to 23.24, 24.91, 24.53, 23.89, and~22.47 in $u^*$, $g^*$,
$r^*$, $i^*$, and $z^*$, respectively.

\noindent
$^b$ Coordinate equinox is J2000.

\bigskip\bigskip\bigskip\bigskip
\halign{\hskip 12pt
# \hfil \tabskip=1em plus1em minus1em&
\hfil # \hfil &
\hfil # \hfil &
\hfil # \hfil \cr
\multispan4{\hfil TABLE 2. Properties of Quasars$^a$ \hfil}\cr
\noalign{\bigskip\hrule\smallskip\hrule\medskip}
\hfil Object \hfil&\hfil $z$ \hfil
&\hfil $AB_{1450}$ \hfil & \hfil $M_B^{\ a}$ \hfil \cr
\noalign{\medskip\hrule\bigskip}
SDSSp J143952.58$-$003359.2   & 4.255 $\pm$ 0.010 & 20.67 & $-24.4$ \cr
SDSSp J143951.60$-$003429.2 & 4.258 $\pm$ 0.010 & 22.03 & $-25.8$ \cr
\noalign{\medskip\hrule}}

\medskip\noindent
$^a$ Calculated assuming $H_0$ = 50, $q_0$ = 0.5, and $\alpha$ = $-0.5$,
and standard Galactic reddening law with $E(B-V)$~=~0.039 (\cite{sfd98}).

\end{document}